# Discovery and construction of surface kagome electronic states induced by *p-d* electronic hybridization

Li Huang[1,2]†, Xianghua Kong[3,4,5]†, Qi Zheng[1,2]†, Yuqing Xing[1,2]†, Hui Chen[1,2], Yan Li[1,2], Zhixin Hu[6], Shiyu Zhu[1,2], Jingsi Qiao[7,4], Yu-Yang Zhang[2], Haixia Cheng[4], Zhihai Cheng[4], Xianggang Qiu[1,2], Enke Liu[1,2], Hechang Lei[4], Xiao Lin[2], Ziqiang Wang[8], Haitao Yang[1,2]*, Wei Ji[4]*, Hong-Jun Gao[1,2,9]*

**Affiliations:**

[1]Beijing National Center for Condensed Matter Physics and Institute of Physics, Chinese Academy of Sciences, Beijing 100190, China

[2]School of Physical Sciences, University of Chinese Academy of Sciences, Beijing 100190, China

[3]College of Physics and Optoelectronic Engineering, Shenzhen University, Shenzhen 518060, China

[4]Beijing Key Laboratory of Optoelectronic Functional Materials & Micro-Nano Devices, Department of Physics, Renmin University of China, Beijing 100872, China

[5]Centre for the Physics of Materials and Department of Physics, McGill University, Montreal QC H3A 2T8, Canada

[6]Center for Joint Quantum Studies and Department of Physics, Institute of Science, Tianjin University, Tianjin 300350, China

[7]MIIT Key Laboratory for Low-Dimensional Quantum Structure and Devices, School of Integrated Circuits and Electronics, Beijing Institute of Technology, Beijing 100081, China

[8]Department of Physics, Boston College, Chestnut Hill, MA, USA

[9]Songshan Lake Materials Laboratory, Dongguan, Guangdong 523808, China

\* Corresponding author. Email: hjgao@iphy.ac.cn, wji@ruc.edu.cn, htyang@iphy.ac.cn

† These authors contributed equally to this work

**Abstract:** Kagome-lattice materials possess attractive properties for quantum computing applications, but their synthesis remains challenging. Herein, we show surface kagome electronic states (SKESs) on a Sn-terminated triangular $Co_3Sn_2S_2$ surface, which are imprinted by vertical *p-d* electronic hybridization between the surface Sn (subsurface S) atoms and the buried Co kagome lattice network in the $Co_3Sn$ layer under the surface. Owing to the subsequent lateral hybridization of the Sn and S atoms in a corner-sharing manner, the kagome symmetry and topological electronic properties of the $Co_3Sn$ layer is proximate to the Sn surface. The SKESs and both hybridizations were verified via qPlus non-contact atomic force microscopy (nc-AFM) and density functional theory calculations. The construction of SKESs with tunable properties can be achieved by the atomic substitution of surface Sn (subsurface S) with other group III-V elements (Se or Te), which was demonstrated theoretically. This work exhibits the powerful capacity of nc-AFM in characterizing localized topological states and reveals the strategy for synthesis of large-area transition-metal-based kagome lattice materials using conventional surface deposition techniques.

Transition metal (TM)-based kagome materials provide an attractive platform for investigating correlated topological properties [1-15] and developing kagome lattice applications [16]. However, the number of already synthesized kagome materials is limited, thus the construction of tailored kagome electronic bandstructures remains challenging[12]. Kagome physics is generally two-dimensional (2D) but kagome electronic states (KESs) are usually induced by kagome planes embedded in the bulk material. These planes of most known TM-based kagome materials are buried under cleavable surfaces of these crystals[14, 17] because the kagome planes usually exhibit higher surface energies than their vicinal planes do[10], which is a roadblock for direct observation and manipulation of KESs in these materials. This issue, however, can be addressed if we are able to build KESs on a surface of a relatively low surface energy (termed surface kagome electronic states, SKESs). Building the SKESs enables directly real-space observation and manipulation of rich kagome physics in the atomic limit by surface characterization techniques, such as scanning tunneling microscopy (STM) and atomic force microscopy (AFM). As a consequence, feasible tunability of the SKESs could also be achieved by tip manipulation or molecular beam epitaxy.

Introduction of a capping layer composed of nonmetal atoms over a kagome plane often lowers the surface energy of the kagome plane[10]. A SKES is thus built if the capping layer could inherit the underlying kagome symmetry electronically. This strategy is, however, unsuitable for charge-transfer (CT) insulators, e.g. copper oxides, in which the TM $d$ bands and $p$ bands of surface adatoms are separated and independent (Fig. 1a) [18, 19]. Fortunately, most TM-based kagome materials are CT metals, where the $p$ bands of non-metal atoms overlap and hybridize with the TM $d$ bands near the Fermi level (Fig. 1b) [9, 12, 14]. This type of hybridization can potentially electronically imprint the kagome symmetry of a less stable kagome plane onto a more stable but hexagonal or triangular surface plane, as schematically shown in Fig. 1c.

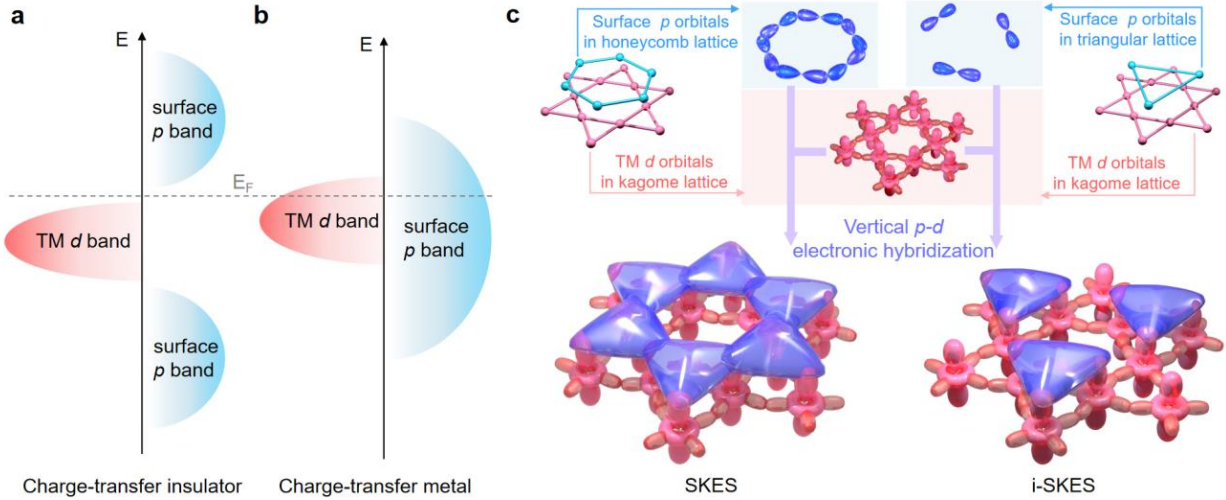

*Fig. 1 | Schematic of vertically hybridized electronic states in charge-transfer kagome metal $Co_3Sn_2S_2$.* Schematic of the energy bands near Fermi level in (**a**) charge-transfer insulator and (**b**) charge-transfer metal, showing the difference in their p-d orbital overlap on the cleaved surfaces. (**c**) Schematic of surface kagome electronic state (SKES, the left part) and incomplete surface kagome electronic state (i-SKES, the right part) formation through vertical p-d hybridization in a charge-transfer kagome metal.

$Co_3Sn_2S_2$, a kagome magnetic Weyl semimetal, is a good candidate to verify whether SKESs could be induced on a non-kagome-latticed surface of a kagome materials. Two major types of its surfaces (Type-I and Type-II) show triangular lattices in scanning tunneling microscopy (STM) images[10, 17, 20-23] but exhibit kagome-related electronic properties[17, 23]. The triangular lattices observed in STM images and a comparison of theoretical surface energies indicate that these two surfaces are, most likely, not $Co_3Sn$-terminated ones, but are covered by S or/and Sn atoms. In line with the strategy proposed in Fig.1c, we infer that surface Sn and subsurface S atoms, which forms a corrugated honeycomb lattice on the Sn-terminated surface, could electronically imprint the kagome symmetry of the $Co_3Sn$ layer forming a SKES (the left part in Fig. 1c). On the S-terminated surface, surface S atoms might have chance to inherit the triangular symmetry of the $Co_3$ trimer underneath (the right part in Fig. 1c). This state could be regarded as an incomplete SKES, in which negligible electron hopping exists in one of the two kagome triangles, as termed incomplete-SKES (i-SKES).

Herein, we experimentally and theoretically verified our inference for building SKESs on $Co_3Sn_2S_2$ surfaces and theoretically tested to tune the SKES by substitute transition metal atoms for surface Sn and S atoms. SKESs (i-SKESs) were observed in non-contact atomic force microscopy (nc-AFM) images on the Type-II (-I) surface. Force spectra and associated density functional theory (DFT) calculations were used to confirm the assignment of type-II (-I) surface to the Sn- (S-) terminated surface. The nc-AFM images were interpreted with the help of DFT calculations, which reveal that the strong *p-d* hybridization of the *p* orbitals of surface Sn or subsurface S with the *d* orbitals of the $Co_3$ kagome network underneath occurs near the Fermi level. The hybridization leads to an SKES (i-SKES) imprinted on the Sn-terminated Type-II (S-terminated Type-I) surface. We also theoretically generalized this SKES construction strategy by substituting the surface Sn (S) atoms with group III-A, IV-A, or V-A element (Se or Te) atoms.

**SKES on Type-II surface of $Co_3Sn_2S_2$**

The bulk crystal of $Co_3Sn_2S_2$ belongs to space group $R\bar{3}m$, comprising a triangular lattice with constants $a$ = 5.37 Å and $c$ = 13.15 Å[11]. As shown in Fig. 2a and Extended Data Fig. 1, $Co_3Sn_2S_2$ has a layered structure composed of a kagome $Co_3Sn$ plane (red atoms) sandwiched between two triangular S planes (yellow atoms), which are then further encapsulated by two separate triangular Sn planes (blue atoms). Unlike STM images, which provide information on delocalized states, qPlus nc-AFM images could reveal the spatial gradients of short-range repulsive interactions resulting from localized electronic states at the single-chemical-bond level[24-33]. Consequently, it can trace electronic orbitals and/or interactions with unprecedented resolution[24, 34]. Therefore, we performed nc-AFM imaging and short-range force spectroscopy on cleaved $Co_3Sn_2S_2$ surfaces. The AFM was equipped with a qPlus sensor, and a CO-functionalized tip was used for imaging (Fig. 2b).

There are two commonly obtained cleaved $Co_3Sn_2S_2$ surfaces: one consisting of a few vacancies (Type-I) and the other containing many adatoms (Type-II). By local contact potential difference (LCPD) measurement, the Type-I surface, with a much higher work function, is demonstrated most likely to be the S surface[23]. However, the $Co_3Sn$ and Sn surfaces have comparable surface work functions and cannot be convincingly distinguished by LCPD, thus leaving the Type-II surface identification a remaining issue. Figure 2c shows an STM image of the S (Type-I) surface, which features a triangular lattice with a few S vacancies. In the nc-AFM image (Fig. 2d), the surface S atoms appeared as bright spots with slightly anisotropy (region $α_I$), maintaining their triangular pattern. A further zoomed-in image shown in Fig. 2e shows three line-like features, equivalently distributed in terms of the rotational angle, extending from each S atom. These "lines" converge to rather blurry regions (the $β_I$ region) in the image, while the remaining dark region is denoted as $γ_I$, implying the formation of i-SKES.

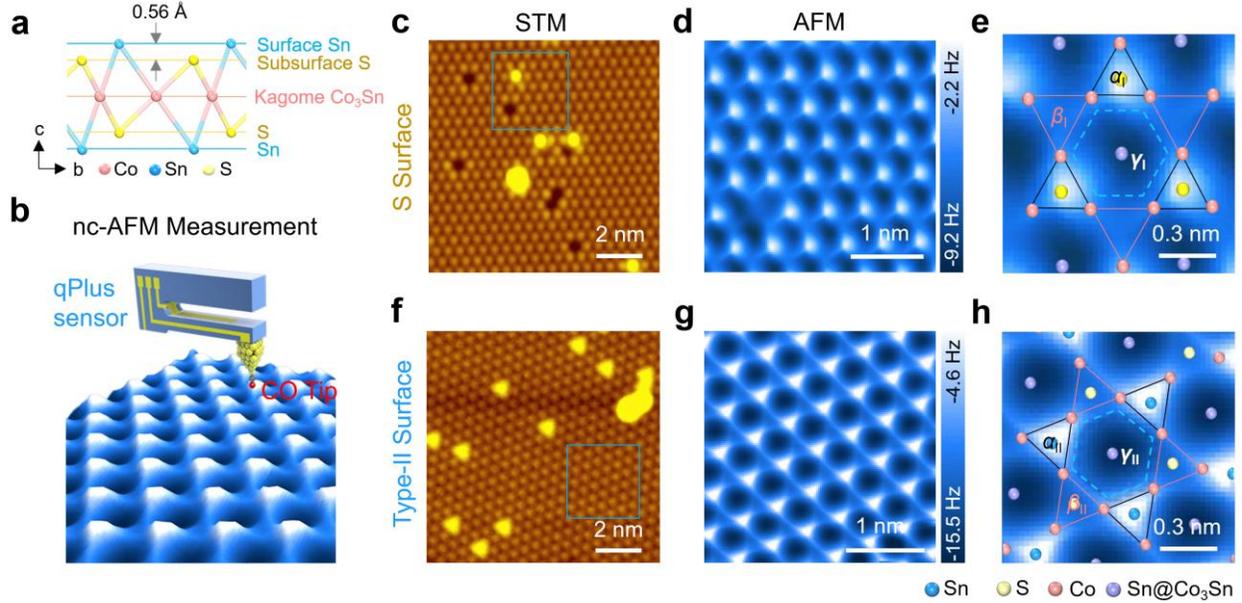

*Fig. 2 | Surface kagome electronic structure of the Sn surface in $Co_3Sn_2S_2$. (a)* Side view of the atomic model of the vertically stacked Sn-S-$Co_3$Sn-S-Sn layers in $Co_3Sn_2S_2$. *(c)* STM image of the S surface of $Co_3Sn_2S_2$. *(b)* Schematic of the nc-AFM measurements using a qPlus sensor with a CO-functionalized tip in the frequency modulation mode. *(d)* Chemical-bond-resolved nc-AFM image of the S surface taken in the blue square in (c). *(e)* Zoomed-in image from (d) to show the incomplete kagome lattice. Three distinct regions within a unit cell with bright, blurry, and dark contrast, which are marked by black solid line triangles, red solid line triangles, and blue dashed line hexagon, are labeled as $α_I$, $β_I$, and $γ_I$ regions. The atomic structure of S surface with the underlying $Co_3$Sn plane is superimposed. *(f)* STM image of the Sn surface of $Co_3Sn_2S_2$. *(g)* Chemical-bond-resolved nc-AFM image of the Sn surface taken in the area marked by a blue square in (f). *(h)* Zoomed-in image from (g), showing the kagome lattice. Three distinct regions within a unit cell with bright, blurry, and dark contrast, which are marked by black solid line triangles, red solid line triangles, and blue dashed line hexagon, are labeled as $α_{II}$, $β_{II}$, and $γ_{II}$ regions. The atomic structure superimposed is the Sn surface with the underlying S and $Co_3$Sn plane. Scanning parameters: (c) and (f), $V_s$ = -400 mV, $I_t$ = 100 pA; (d) and (e), amplitude = 100 pm, scanning height = 180 pm lower from a tunneling junction of $V_s$ = -4 mV, $I_t$ = 10 pA; (g) and

*(h), amplitude = 100 pm, scanning height = 220 pm lower from a tunneling junction of $V_s$ = -400 mV, $I_t$ =100 pA.*

The Type-II surface, which displays a triangular lattice network decorated with a few adatoms in the STM image (Fig. 2f), shows bright dots assembling in the same triangular lattice in the nc-AFM images acquired at relatively large tip-sample distances (Extended Data Fig. 2a). In a zoomed-in nc-AFM image, we found the dots are circular in shape (region $α_{II}$ marked in Extended Data Fig. 2b), which outline a blurry triangular ($β_{II}$), and an indistinctly hexagonal region ($γ_{II}$). As the tip approaches the sample surface (Extended Data Fig. 2c and Extended Data Fig. 2d), the circular shape of the dots evolves into explicitly triangular shape in the $α_{II}$ region and the blurry triangles become sharper in region $β_{II}$. They eventually connect to form a 2D pattern showing a breathing kagome pattern (Fig. 2g). In a zoomed-in image (Fig. 2h), the $α_{II}$ region appear as bright triangles with each side 2.77 Å in length, while the $β_{II}$ region appear bigger but as slightly blurry triangles with each side 3.02 Å in length, which is 9% longer than that of the $α_{II}$ region. The kagome-latticed Type-II surface appears to resemble the structure of the $Co_3Sn$ layer, which is, however, less stable compared to the Sn layer under exposure to the surface[10]. If the kagome-appeared pattern was, as we inferred in Fig. 1c, imprinted electronically from the $Co_3Sn$ layer, the Type-II surface would be the energetically preferred Sn-terminated surface. Force spectra measurements and DFT calculations were thus called to solve this surface identification issue.

**Identification of Type-II surface**

We first measured vertical short-range force spectra on the less controversy S (Type-I) surface to verify the reliability of our experiment and theory comparison. As shown in Fig. 3a, the force spectra at the $α_I$, $β_I$, and $γ_I$ regions all hit the attractive-repulsive tuning point below 3Å, with the respective height of the turning point at the $α_I$ ($γ_I$) regions 0.9 Å higher (0.6 Å lower) than that at the $β_I$ region. In Fig. 3b, we tentatively assessed those three regions to the top-layer S (black dashed circle), the $Co_3$ trimer (red dashed triangle) and the Sn atom (blue dashed triangle) in the $Co_3Sn$ layer, respectively. Given that, our DFT calculations well reproduced those experimental force curves in Fig. 3c, which were plotted using the same color code and exhibit comparable respective heights of the turning positions, i.e., 0.9 Å versus 1.0 Å (0.6 Å versus 0.6 Å) for the $α_I$ ($γ_I$) region, compellingly supporting those tentative assignments. More importantly, these experiment-theory coincidences, in turn, confirm the assessment of the Type-I surface to the S surface and validate our force-spectra-comparison method of identifying surface terminations of $Co_3Sn_2S_2$.

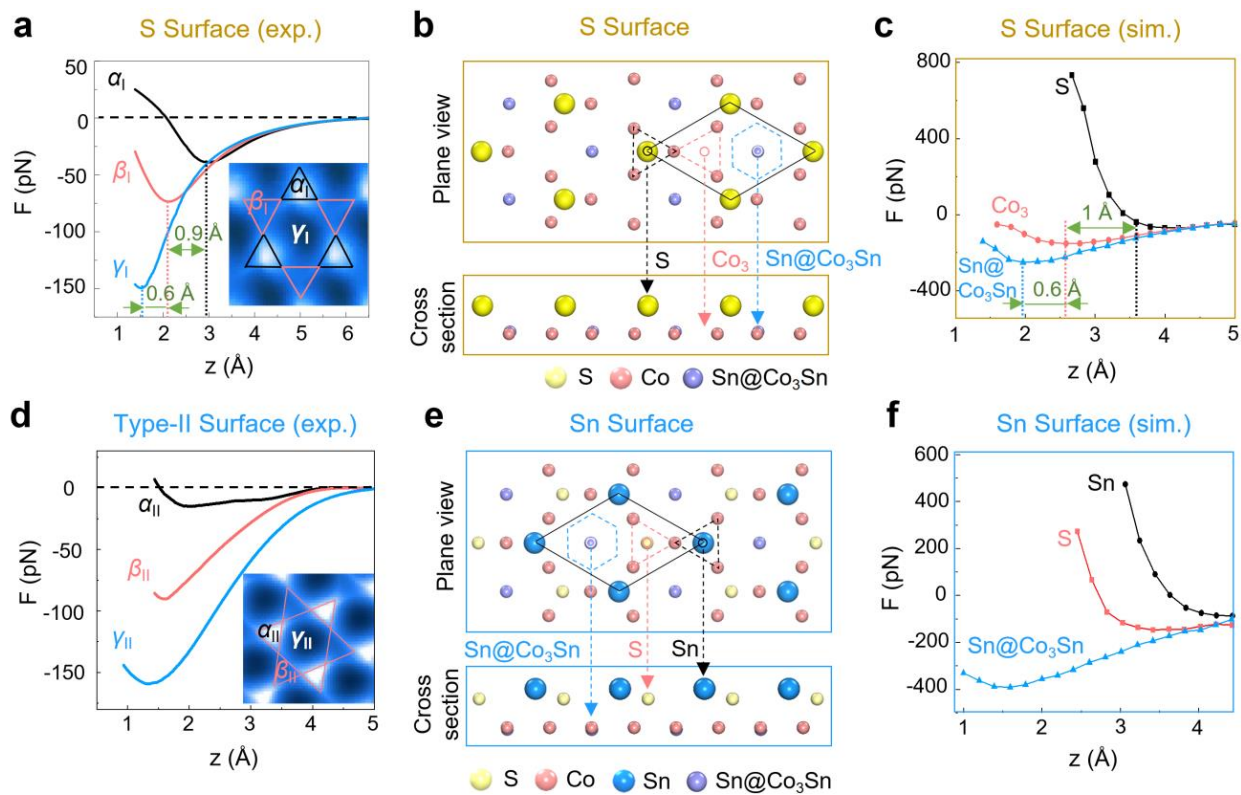

**Fig. 3 | Identification of the Sn surface by short-range force spectra.** (*a*) Experimental vertical short-range force spectra measured at the center of the $α_I$ (black), $β_I$ (red), and $γ_I$ (blue) regions. (*b*) and (*c*) DFT optimized surface structure and vertical short-range force spectra on the S surface. (*d*) Experimental vertical short-range force spectra measured at the center of the $α_{II}$ (black), $β_{II}$ (red), and $γ_{II}$ (blue) regions. (*e*) and (*f*) DFT optimized surface structure and vertical short-range force spectra on the Sn surface. Spectra displayed in (*a*) were obtained with an amplitude of 25 pm and 50 pm for that shown in (*d*). Both experimental vertical short-range force curves were deconvoluted from associated frequency-shift curves using the Sader-Jarvis method[35].

Figure 3d plots the experimental force spectra on the Type-II surface, in the same color and spatial schemes used on the Type-I surface. As the tip approaching the surface, the $α_{II}$ region (black) appears the least attractive among all the three regions and first reaches the turning point at 1.92 Å, while that value is 1.60 Å and 1.31 Å for regions $β_{II}$, and $γ_{II}$, respectively, in which region $γ_{II}$ shows a larger attractive force. We modeled the acquiring process of those force curves on both the Co$_3$Sn and Sn surfaces using DFT calculations. The surface Sn atoms show appreciable surface relaxations on the Co$_3$Sn surface and are 0.42 Å lifted from the Co$_3$Sn plane (the cross section in Extended Data Fig. 3a). Consequently, the force spectrum on the Sn atom (blue curve in Extended Data Fig. 3b) first reaches its minimum and shows strong repulsion at shorter distances. The two triangular regions, i.e., the surface Co$_3$ trimer sit on a Sn atom and a S atom underneath (Co$_3$/Sn and Co$_3$/S) are less repulsive in comparison with the surface Sn (hexagonal) region, apparently inconsistent with the experimental spectra on the Type-II surface where the two triangular regions exhibit stronger repulsion. This contradiction does not support the assessment of the Co$_3$Sn surface to the Type-II surface.

The Sn surface contains triangularly distributed Sn atoms (blue balls in Fig. 3e) in the topmost layer while the same lattice of S atoms (yellow balls), with a lateral shift of (1/3, 1/3) unit cell, is placed in a layer just 0.56 Å below. Each Sn or S atom sits over a Co$_3$ trimer, which could be regarded as a region of triangular symmetry, while the Sn atom in the Co$_3$Sn layer appears a hexagonal symmetry. Thus, we could tentatively assign the Sn and S triangles to triangular regions $\alpha_{II}$ (black) and $\beta_{II}$ (red), respectively, and the Sn@Co$_3$Sn hexagon to hexagonal region $\gamma_{II}$ (blue). Associated theoretical force spectra (Fig. 3f) show consistent results with the experimental ones in terms of the order of repulsion among those three regions. This verifies our assignment that region $\alpha_{II}$ ($\beta_{II}$ and $\gamma_{II}$) represents the Sn (S and Sn@Co$_3$Sn) site, as denoted by the superimposed atomic structure in Fig. 2h. Thereby, based on site-dependent force spectra measurements and corresponding DFT calculations, we identified that Type-II surface, which hosts SKES, is the Sn surface of Co$_3$Sn$_2$S$_2$.

***p-d* hybridization between surface atoms and the kagome plane**

To elucidate the origin of the kagome-shaped feature on the Sn surface revealed in the nc-AFM images in Figs. 2g and 2h, the projected local density of states (PLDOS) for the Sn-terminated Co$_3$Sn$_2$S$_2$ surface is plotted in Fig. 4a. It shows a group of electronic states around the Fermi level ($E_F$), which is nearly isolated from other states and ranges from -0.38 to 0.50 eV, as highlighted by the parallel black dashed lines. These states do not consist of the *s*-orbital component of the surface Sn or S atoms (see Extended Data Fig. 5 for details). We replotted PLDOSs of the *p* orbitals of surface Sn and subsurface S atoms and the *d* orbitals of the Co$_3$ trimers underneath between -0.38 and 0.50 eV in Fig. 4b, which indicate strong electronic hybridization between the surface Sn

(S) $p$ orbitals, denoted in blue (green), and the $d$ states of the underlying $Co_3$ trimers (in orange). The positions of the four pronounced hybridization peaks are highlighted using four dashed lines.

The $p$-$d$ hybridized states effectively impose the triangular-shaped electron density of the triangular $Co_3$ trimers on those of the surface Sn and S atoms. Figure 4c depicts the square of the wavefunction norms ($|\psi|^2$) of the occupied hybridized states, showing a triangular-shaped contour at each Sn or S site. Six of these triangles surround a hexagonal area around the Sn@$Co_3$Sn site, where the contour density is the lowest, forming a kagome lattice network on the surface. The plot of the $|\psi|^2$ contour also reveals the lateral interconnections of the triangularly shaped density contours, indicating their lateral electronic hybridization in real space. The PLDOS confirms this lateral hybridization between Sn and S, which forms electronic states at -0.25 and 0.37 eV (red dashed lines in Fig. 4b). Therefore, through vertical $p$-$d$ hybridization and subsequent lateral $p$-$p$ hybridization, the kagome symmetry of the $Co_3$Sn plane underneath was electronically imprinted on the Sn-terminated surface, and an SKES was successfully demonstrated.

The hybridization of Sn and S with the Co kagome lattice network also transferred the nontrivial properties of the kagome electronic states onto the Sn surface. A scanning tunneling spectral (STS) measurement of the Sn surface (Extended Data Fig. 6a) exhibits a sharp peak at approximately -10 meV ($P_k$), as indicated by the red arrow. This sharp feature fits well with the two flat bands near $E_F$ of the Sn surface (Extended Data Fig. 6b), which are SKESs that are laterally hybridized between Sn-$Co_3$ and S-$Co_3$ states. The position of $P_k$ shifts toward $E_F$ under an applied magnetic field perpendicular to the surface with both up- and down-field orientations (Extended Data Fig. 6c). Such an unconventional splitting trend indicates negative flat band magnetism, further confirming the topological origin of the lateral hybridization SKESs.

DFT calculation results for the S surface are displayed in Fig. 4d-4f for comparison with those of the Sn surface. The associated PLDOS spectra were plotted in Fig. 4d. It exhibits a group of isolated states between 0.12 and 1.16 eV, where only the $p$ orbitals of S (both at surface and in bulk) and $d$ orbitals of Co are involved. In this energy window, a zoomed-in PLDOS plot of the surface S and subsurface $Co_3$Sn layers, also indicates vertical hybridization between the surface S $p$ states and the underlayer Co $d$ states, denoted by the dashed lines in Fig. 4e. These vertical $p$-$d$ hybridizations impose triangular-shaped $Co_3$ electronic states on the surface S atoms, as revealed by the $|\psi|^2$ contour in Fig. 4f. These hybridized states on S atoms could, as we illustrated in Fig. 1d and Fig. 2e, be regarded as an i-SKES because there are no electronic states on top of the other sublattice over the Co kagome network (marked by red solid line triangles in Fig. 4f). As a result, the flat bands observed around $E_F$ on the Sn surface were absent on the S surface, even in a wider energy window [10, 23].

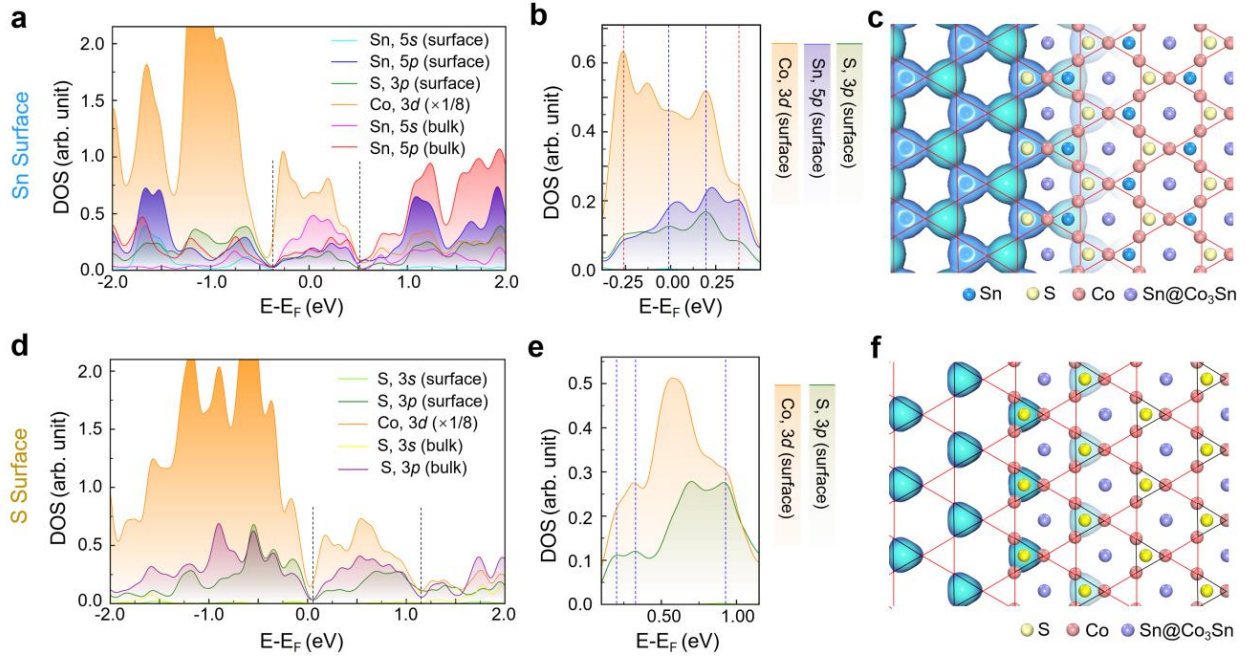

*Fig. 4 | Vertical p-d hybridizations at the Sn and S surfaces in $Co_3Sn_2S_2$. (a)* PLDOS of the Sn surface. Two black dashed lines at -0.38 and 0.50 eV indicate the boundaries of a group of isolated p-d hybridized states. *(b)* PLDOS of the top three atomic layers of the Sn surface, from -0.38 to 0.50 eV. Blue and red dashed lines denote the vertically p-d hybridized states between the surface (subsurface) Sn (S) atoms and the $Co_3$ trimers underneath, while the red dashed lines also indicate in-plane (laterally) p-p hybridized states between surface Sn and subsurface S atoms. *(c)* Isosurface contour of $|\psi|^2$, integrated from -0.38 eV to the $E_F$, on the Sn surface superposed with the atomic structure of the top three atomic layers of the surface in the right part. Red solid lines highlight the kagome lattice. *(d-f)* Duplicate PLDOS and $|\psi|^2$ isosurface contour plots, in the same schemes used in *(a-c)*, for the S surface. Boundaries of the isolated p-d hybridized states is at 0.12 and 1.16 eV, as marked by the two black dashed lines in *(d)*. *(e)* Zoomed-in PLDOS plotted from 0.12 to 1.16 eV. *(f)* Isosurface contour of $|\psi|^2$ integrated from 0.12 to 1.16 eV. The black solid lines mark the hybridized states on S atoms; while the red solid lines mark the missing sublattice of the i-SKES.

**Strategy for constructing a family of SKESs**

We have experimentally and theoretically demonstrated the feasibility of SKES construction. The construction strategy requires that (i) the surface and subsurface (if any) atoms fit in a honeycomb lattice, (ii) their in-plane states vertically hybridize with both of the corner-sharing triangular-shaped sublattices of the kagome lattice underneath, and (iii) their hybridized states then subsequently hybridize laterally with each other to form an SKES, as illustrated in Fig. 5a. In $Co_3Sn_2S_2$, the Sn terminated surface approximately meets all these requirements, in which the surface Sn atoms are located on one sublattice, and subsurface S atoms, only 0.56 Å below the surface, reside on the other sublattice. The S-terminated surface does not meet requirement (i), but the additional deposition of hetero- or homo-atoms onto the other sublattice (sublattice plane **a**, SLa, see Fig. 5b) fills the incomplete part of the i-SKES, enabling the construction of SKESs on the S surface with tunable properties. Furthermore, we could obtain more diverse types of SKESs by substituting the S atoms with Se or Te (sublattice plane **b**, SLb, see Fig. 5b) during bulk crystal growth.

DFT calculations were performed to examine the geometric and electronic structures of the surfaces with modified SLa and SLb planes. Diverse surface electronic hybridizations and SKESs were verified. Similar vertical $p$-$d$ electronic hybridizations are preserved when SLa = Ge, Sn, or Pb and SLb = S or Se. A typical $|\psi|^2$ contour of the isolated states near $E_F$ for SLa = Sn and SLb = Se was shown in Fig. 5c, revealing kagome features that are comparable to those of the Sn-terminated $Co_3Sn_2S_2$ surface. The PLDOS (Extended Data Fig. 7a) indicate that the hybridization states comprising Sn and Se $p$ orbitals and $d$ orbitals of Co are restricted within the energy range of -0.5 to 0.5 eV. They are isolated from other electronic states and can therefore be considered pure $p$-$d$ hybridized SKES near $E_F$. We also constructed a more typical breathing SKES by more significantly unbalancing the hybridized states distributed on the SLa and SLb sites, as achieved by substituting the SLa site with heavier elements, like Sb and Bi (Fig. 5d and Extended Data Fig. 7b). For SLa = Al, Ga, In, or Si or SLb = Te, their $s$ orbitals, in addition to the $p$-$d$ hybridization, are involved in vertical hybridization with Co around $E_F$ (Fig. 5e and Extended Data Fig. 7c). The involvement of the $s$ orbital results in the $|\psi|^2$ contour of the hybridized states appearing spherical in shape, which significantly weakens the kagome feature on the surface. The energetically isolated hybridized states are mixed with other states for surfaces where SLa = P or As, in which the surface kagome feature is eliminated (Extended Data Fig. 7d). These results reveal that the S surface of $Co_3Sn_2S_2$ can be used as a template for constructing various SKESs with tailored properties.

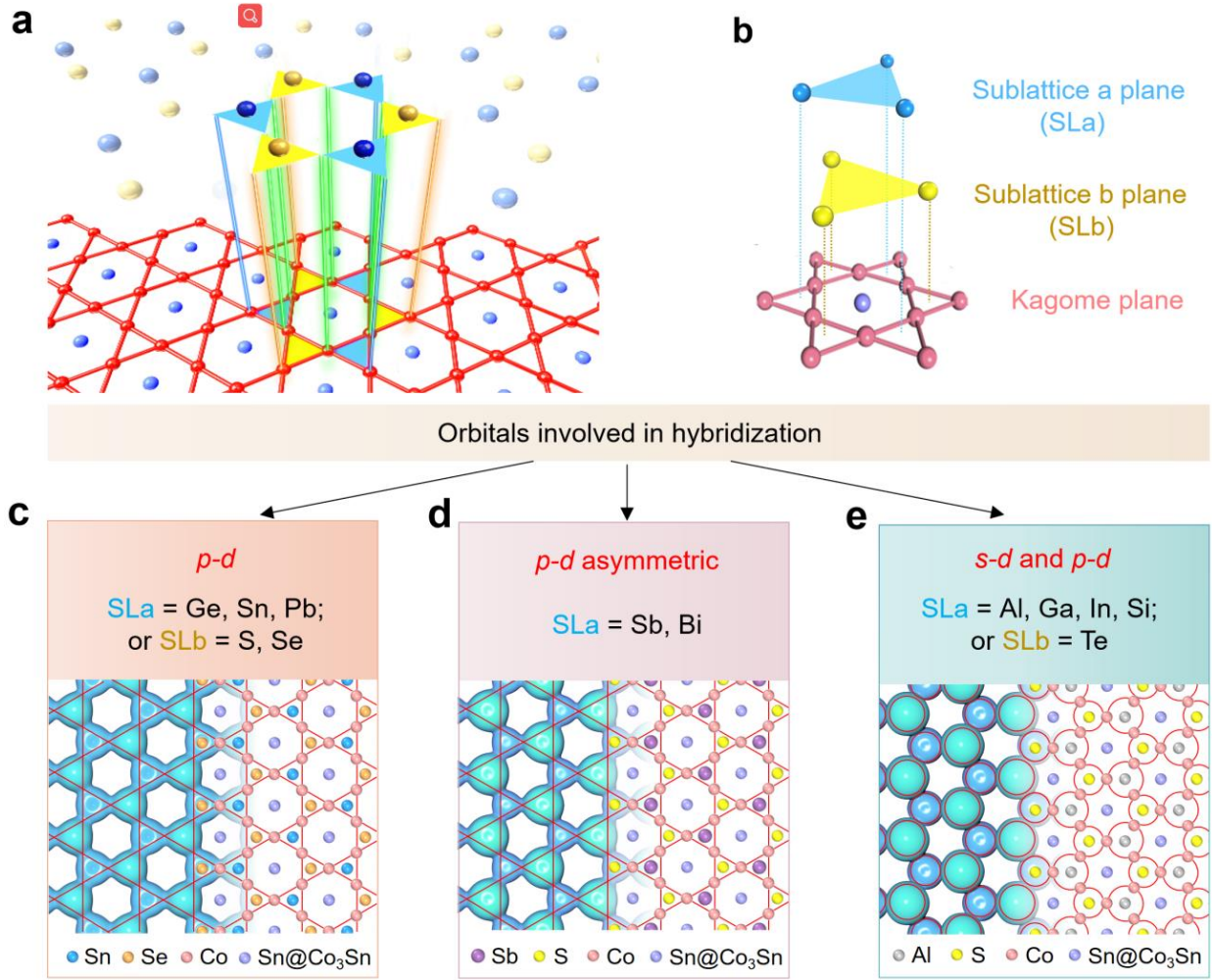

*Fig. 5 | Theoretical strategy for artificially constructing a family of surface kagome electronic states. (**a**) Schematic of the vertical p-d hybridization between the $Co_3$ trimer and the surface atoms. Blue and yellow triangles indicate the hybridized triangular-shaped electronic state on the two sublattices of the kagome symmetry, respectively. (**b**) Schematic of the surface planes located on top of sublattice **a** (blue, SLa plane) and sublattice **b** (yellow, SLb plane) of the kagome plane (red). (**c-e**) Plots of $|\psi|^2$ isosurface contours of hybridized states. (**c**) Contour of a p-d hybridized SKES for SLa = Sn and SLb = Se. Deposition of Ge and Pb on the S surface also generate the same type of SKESs. (**d**) Contour of a p-d hybridized asymmetric SKES for SLa = Sb and SLb = S, where the S and Sb sites show substantially different intensities. Deposition of Bi on the S surface also generate the same type of SKESs. (**e**) Contour for SLa = Al and SLb = S, where*

*both s-d and p-d mixed hybridizations are involved. The same type of SKESs can be constructed with deposition of Ga, In and Si on S surface, or substitution of SLb elements with Te.*

**Conclusions**

In summary, we have demonstrated a strategy for constructing energetically stable SKESs that involves electronically imprinting the kagome lattice symmetry of an undersurface layer through vertical *p-d* and lateral *p-p* electronic hybridizations. The strategy was verified using the Sn surface of $Co_3Sn_2S_2$, which is a prototypical kagome magnetic Weyl semimetal. By combining chemical-bond-resolved nc-AFM images, vertical short-range force spectra, and DFT calculations, we explicitly found SKESs on the Sn surface of $Co_3Sn_2S_2$. Our DFT calculations revealed that these states originate from the strong vertical electronic hybridization between triangularly arranged surface Sn or subsurface S atoms and the $Co_3$ trimers of the $Co_3Sn$ kagome lattice underneath, accompanied by lateral hybridizations of the surface Sn and subsurface S states. STS measurements verified the appearance of kagome symmetry-featured topological properties on the Sn surface. To generalize this strategy, DFT calculations explored the surface electronic structure variations for atomic substitutions of surface Sn (subsurface S) with III-A, V-A, and IV-A (Se or Te) elements on the Sn surface. We also theoretically demonstrated the feasibility of constructing surface kagome structures by depositing group III-A or IV-A elements on the S surface. The demonstration of nc-AFM for investigating localized electronic surface states aside, our strategy for building SKES enriches the design routes to construct semi-metallic 2D kagome materials with topologically non-trivial properties.

**Methods**

*Single crystal growth of $Co_3Sn_2S_2$*

A Sn/Pb mixed flux was used to grow single crystals of $Co_3Sn_2S_2$. First, the starting materials were mixed at a molar ratio of Co:S:Sn:Pb = 12:8:35:45 (Co (99.95% Alfa), Sn (99.999% Alfa), S (99.999% Alfa), and Pb (99.999% Alfa)). The mixture was placed in an $Al_2O_3$ crucible sealed in a quartz tube that was slowly heated to 673 K for 6 h and then maintained for 6 h to avoid heavy loss of sulfur. Thereafter, the quartz tube was heated to 1323 K for 6 h and maintained for 6 h before slowly cooling to 973 K over 70 h. At 973 K, rapid decanting and subsequent spinning in a centrifuge were performed to remove flux. Finally, hexagonal-plate single crystals with diameters of 2-5 mm were obtained. Energy-dispersive X-ray spectroscopy and X-ray diffraction were used to determine the compositions and phase structures of the crystals.

*QPlus nc-AFM measurements*

Non-contact AFM measurements were performed on a combined nc-AFM/STM system (Createc) at 4.7 Kwith a base pressure lower than $2 \times 10^{-10}$ mbar. All measurements were performed using a commercial qPlus tuning fork sensor in the frequency modulation mode with a Pt/Ir tip at 4.5 K. The resonance frequency of the AFM tuning fork was 27.9 kHz, and the stiffness was approximately 1800 N/m. The $Co_3Sn_2S_2$ samples were cleaved at <10 K in an ultrahigh vacuum chamber and transferred to the nc-AFM/STM head within 10 s. nc-AFM images and short-range force spectra were recorded using CO-functionalized tips. The oscillation amplitudes used in all measurements are stated in the corresponding figure captions.

*DFT calculations*

DFT calculations were performed using the generalized gradient approximation for the exchange-correlation potential, the projector augmented wave method [36] and a plane-wave basis set as implemented in the Vienna ab-initio simulation package [37, 38]. The energy cutoff for the plane-wave basis was set to 400 eV for structural relaxations and 500 eV for the energy and electronic structure calculations. Two $k$-meshes of $7 \times 7 \times 1$ and $11 \times 11 \times 1$ were adopted for the structural relaxations and total energy (electronic structure) calculations, respectively. The mesh density of the k points was fixed when performing the related calculations with primitive cells. In geometric structure relaxation, van der Waals (vdW) interactions were considered at the vdW-DF level with the optB86b functional as the exchange functional (optB86b-vdW) [39, 40]. Symmetrical slab models were employed, and the surface atoms were fully relaxed until the residual force per atom was less than 0.005 eV·Å$^{-1}$. To avoid image interactions between adjacent unit cells, a vacuum layer of more than 20 Å thick was added to the slab cell perpendicular to the surface. The optimized lattice constants of bulk $Co_3Sn_2S_2$ are 5.37 and 13.15 Å along the *a* and *c* directions, respectively. A tip-sample model was established for the force spectral calculations. In particular, the AFM tip was modeled using a five-layer thick Pt(111) cluster, the bottom of which adsorbs a CO molecule, and a (4 × 4) $Co_3Sn_2S_2$ slab was employed to model the sample surface. Our tip model exhibited a significant p-wave feature, which contributes to the high resolution in images acquired using CO-functionalized tips [41, 42]. Energetically favorable magnetic ground states were calculated and used for all the tip-sample configurations.

## References

1. D. L. Bergman, C. Wu, L. Balents, Band touching from real-space topology in frustrated hopping models. *Phys. Rev. B* **78**, 125104 (2008).
2. H. M. Guo, M. Franz, Topological insulator on the kagome lattice. *Phys. Rev. B* **80**, 113102 (2009).
3. I. I. Mazin *et al.*, Theoretical prediction of a strongly correlated Dirac metal. *Nat. Commun.* **5**, 4261 (2014).
4. S. Nakatsuji, N. Kiyohara, T. Higo, Large anomalous Hall effect in a non-collinear antiferromagnet at room temperature. *Nature* **527**, 212-215 (2015).
5. K. Kuroda *et al.*, Evidence for magnetic Weyl fermions in a correlated metal. *Nat. Mater.* **16**, 1090-1095 (2017).
6. L. Ye *et al.*, Massive Dirac fermions in a ferromagnetic kagome metal. *Nature* **555**, 638-642 (2018).
7. Z. Lin *et al.*, Flatbands and Emergent Ferromagnetic Ordering in $Fe_3Sn_2$ Kagome Lattices. *Phys. Rev. Lett.* **121**, 096401 (2018).
8. Q. Wang *et al.*, Large intrinsic anomalous Hall effect in half-metallic ferromagnet $Co_3Sn_2S_2$ with magnetic Weyl fermions. *Nat. Commun.* **9**, 3681 (2018).
9. E. Liu *et al.*, Giant anomalous Hall effect in a ferromagnetic kagome-lattice semimetal. *Nat. Phys.* **14**, 1125-1131 (2018).
10. N. Morali *et al.*, Fermi-arc diversity on surface terminations of the magnetic Weyl semimetal $Co_3Sn_2S_2$ *Science* **365**, 1286-1291 (2019).
11. D. F. Liu *et al.*, Magnetic Weyl semimetal phase in a Kagome crystal. *Science* **365**, 1282-1285 (2019).
12. M. Kang *et al.*, Dirac fermions and flat bands in the ideal kagome metal FeSn. *Nat. Mater.* **19**, 163-169 (2020).
13. J.-X. Yin *et al.*, Quantum-limit Chern topological magnetism in $TbMn_6Sn_6$. *Nature* **583**, 533-536 (2020).
14. H. Chen *et al.*, Roton pair density wave in a strong-coupling kagome superconductor. *Nature* **599**, 222-228 (2021).
15. T. Neupert, M. M. Denner, J.-X. Yin, R. Thomale, M. Z. Hasan, Charge order and superconductivity in kagome materials. *Nat. Phys.* **18**, 137-143 (2022).
16. A. Hutter, J. R. Wootton, D. Loss, Parafermions in a Kagome Lattice of Qubits for Topological Quantum Computation. *Phys. Rev. X* **5**, 041040 (2015).
17. J.-X. Yin *et al.*, Negative flat band magnetism in a spin–orbit-coupled correlated kagome magnet. *Nat. Phys.* **15**, 443-448 (2019).
18. B. Keimer, S. A. Kivelson, M. R. Norman, S. Uchida, J. Zaanen, From quantum matter to high-temperature superconductivity in copper oxides. *Nature* **518**, 179-186 (2015).
19. Ø. Fischer, M. Kugler, I. Maggio-Aprile, C. Berthod, C. Renner, Scanning tunneling spectroscopy of high-temperature superconductors. *Rev. Mod. Phys.* **79**, 353-419 (2007).

20. L. Jiao *et al.*, Signatures for half-metallicity and nontrivial surface states in the kagome lattice Weyl semimetal $Co_3Sn_2S_2$. *Phys. Rev. B* **99**, 245158 (2019).
21. S. Howard *et al.*, Observation of linearly dispersive edge modes in a magnetic Weyl semimetal $Co_3Sn_2S_2$. arXiv:1910.11205 [cond-mat.mtrl-sci] (2019).
22. J.-X. Yin *et al.*, Spin-orbit quantum impurity in a topological magnet. *Nat. Commun.* **11**, 4415 (2020).
23. Y. Xing *et al.*, Localized spin-orbit polaron in magnetic Weyl semimetal $Co_3Sn_2S_2$. *Nat. Commun.* **11**, 5613 (2020).
24. L. Gross, F. Mohn, N. Moll, P. Liljeroth, G. Meyer, The Chemical Structure of a Molecule Resolved by Atomic Force Microscopy. *Science* **325**, 1110-1114 (2009).
25. L. Gross *et al.*, Bond-Order Discrimination by Atomic Force Microscopy. *Science* **337**, 1326-1329 (2012).
26. M. Emmrich *et al.*, Subatomic resolution force microscopy reveals internal structure and adsorption sites of small iron clusters. *Science* **348**, 308-311 (2015).
27. A. Riss *et al.*, Imaging single-molecule reaction intermediates stabilized by surface dissipation and entropy. *Nat. Chem.* **8**, 678-683 (2016).
28. S. Kawai *et al.*, Superlubricity of graphene nanoribbons on gold surfaces. *Science* **351**, 957-961 (2016).
29. N. Pavliček *et al.*, Synthesis and characterization of triangulene. *Nat Nano* **12**, 308-311 (2017).
30. K. Kaiser *et al.*, An sp-hybridized molecular carbon allotrope, cyclo 18 carbon. *Science* **365**, 1299-1301 (2019).
31. J. Berwanger, S. Polesya, S. Mankovsky, H. Ebert, F. J. Giessibl, Atomically Resolved Chemical Reactivity of Small Fe Clusters. *Phys. Rev. Lett.* **124**, 096001 (2020).
32. J. Qi *et al.*, Force-Activated Isomerization of a Single Molecule. *J. Am. Chem. Soc.* **142**, 10673-10680 (2020).
33. F. Huber *et al.*, Chemical bond formation showing a transition from physisorption to chemisorption. *Science* **366**, 235-238 (2019).
34. A. Liebig, P. Hapala, A. J. Weymouth, F. J. Giessibl, Quantifying the evolution of atomic interaction of a complex surface with a functionalized atomic force microscopy tip. *Sci. Rep.* **10**, 14104 (2020).
35. J. E. Sader, S. P. Jarvis, Accurate formulas for interaction force and energy in frequency modulation force spectroscopy. *Appl. Phys. Lett.* **84**, 1801-1803 (2004).
36. P. E. Blochl, Projector augmented-wave method. *Phys. Rev. B* **50**, 17953-17979 (1994).
37. G. Kresse, J. Furthmuller, Efficiency of ab-initio total energy calculations for metals and semiconductors using a plane-wave basis set. *Computational Materials Science* **6**, 15-50 (1996).
38. G. Kresse, J. Furthmüller, Efficient iterative schemes for ab initio total-energy calculations using a plane-wave basis set. *Phys. Rev. B* **54**, 11169-11186 (1996).
39. J. Klimeš, D. R. Bowler, A. Michaelides, Van der Waals density functionals applied to solids. *Phys. Rev. B* **83**, 195131 (2011).
40. J. Klimeš, D. R. Bowler, A. Michaelides, Chemical accuracy for the van der Waals density functional. *J. Phys. Condens. Matter* **22**, 022201 (2009).
41. L. Gross *et al.*, High-Resolution Molecular Orbital Imaging Using a *p*-Wave STM Tip. *Phys. Rev. Lett.* **107**, 086101 (2011).
42. H. Mönig *et al.*, Quantitative assessment of intermolecular interactions by atomic force microscopy imaging using copper oxide tips. *Nat. Nanotech.* **13**, 371-375 (2018).

**Acknowledgments:**

The work is supported by grants from the National Key Research and Development Projects of China (2019YFA0308500, 2018YFE0202700, and 2018YFA0305800), the National Natural Science Foundation of China (61888102, 61761166009, 11974394, 11974422 and 12104313), the Chinese Academy of Sciences (XDB30000000, YSBR-003, 112111KYSB20160061), the Fundamental Research Funds for the Central Universities of China and the Research Funds of Renmin University of China (22XNKJ30), the Department of Science and Technology of Guangdong Province grant 2021QN02L820 and the Shenzhen Natural Science Fund (the Stable Support Plan Program 20220810161616001). Z.Q.W. is supported by the US DOE, Basic Energy Sciences Grant No. DE-FG02-99ER45747. X.K. thanks Prof. Hong Guo for his financial support to the work done at McGill University.

**Author contributions:**

H.-J.G. and W.J. conceive of this project; L.H., Q.Z., Y.X., H.C., Z.C. and X.L. performed NC-AFM/STM experiments with guidance of H.-J.G.; X.K., Z.H., Y.Z., H.C. and W.J. carried out theoretical calculations and analysis; H.Y., X.Q., E.L., and H.L. prepared the samples; Y.L., S.Z and J.Q. helped in plotting the figures; L.H., Q.Z., X.K., Z.W., W.J. and H.-J.G. write the manuscript with inputs from all authors.

**Competing interests:** Authors declare that they have no competing interests.

**Data and materials availability:** Data measured or analysed during this study are available from the corresponding author on reasonable request.

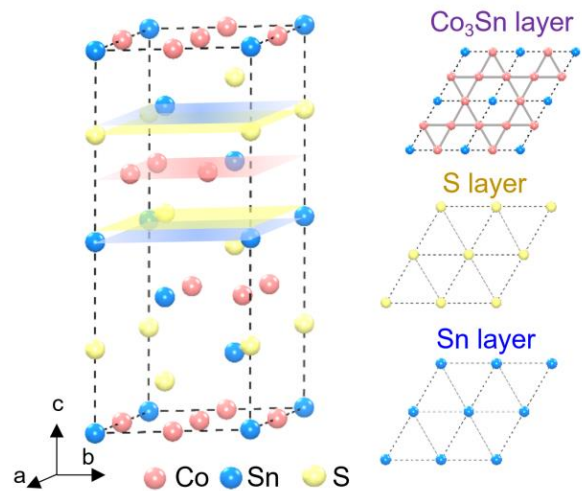

**Extended Data Fig. 1 | Atomic structure of $Co_3Sn_2S_2$.**

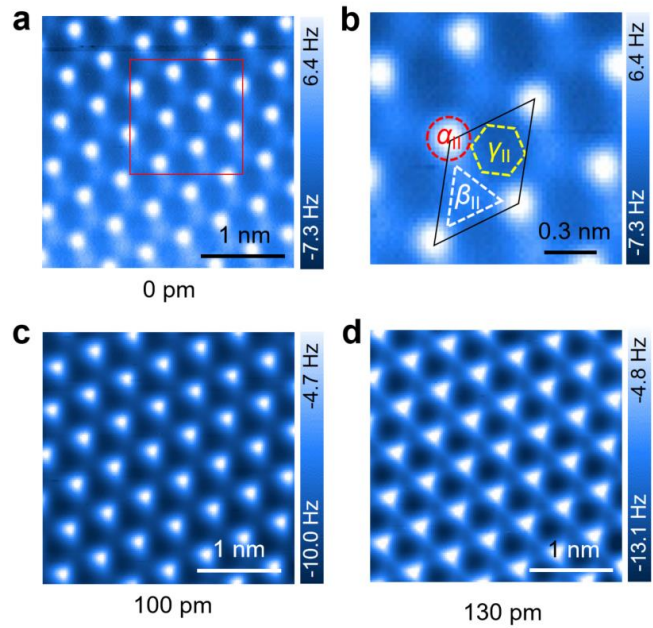

**Extended Data Fig. 2 | nc-AFM images at different scanning heights on Type-II surface.** (**a**) Constant-height nc-AFM image scanned at the tunneling height of $V_s$ = -4 mV, $I_t$=10 pA. (**b**) Enlarged from the red square in (**a**) to show the $α_{II}$, $β_{II}$, and $γ_{II}$ regions within a supercell. (**c**) and (**d**) Constant-height nc-AFM images at different scanning heights on the Type-II surface. The x and y directions have drifted a little during the scanning time. The number below each image is the scanning height the lowered from a tunneling junction height of $V_s$ = -4 mV, $I_t$=10 pA. Scanning amplitude = 50 pm.

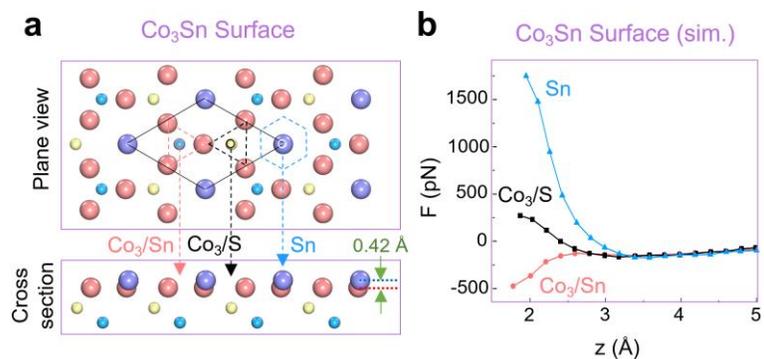

**Extended Data Fig. 3 | DFT calculated vertical short-range force spectrum on optimized Co₃Sn surface.** (**a**) DFT optimized surface structures of the Co₃Sn surface. (**b**) DFT calculated vertical short-range force spectra on the three typical regions marked by the blue hexagon, black and red triangles in (**a**), plotted with the same color code.

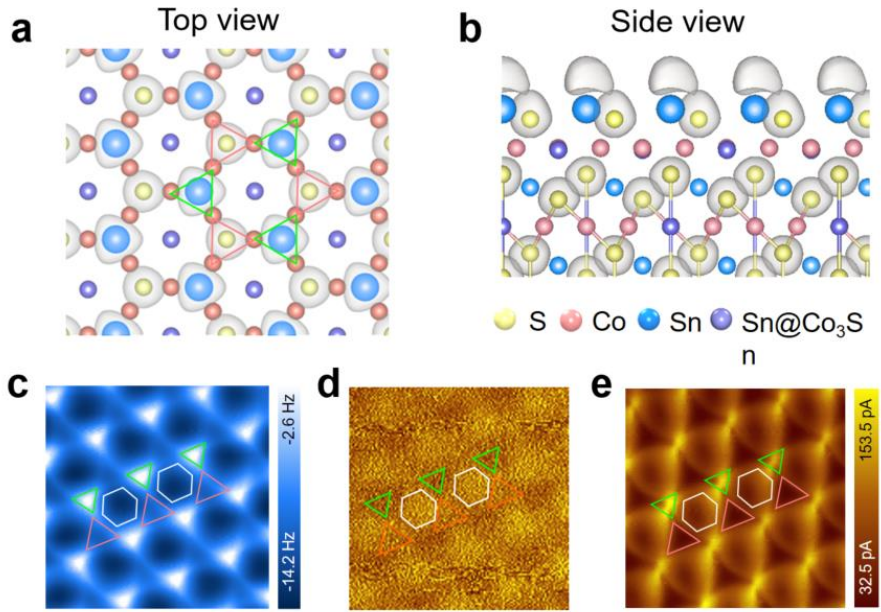

**Extended Data Fig. 4 | Localized electronic states of the surface Sn atoms through *p-d* hybridization on Sn surface.** (**a**) and (**b**) Top and side view of electron localized function (ELF). Both the surface Sn and subsurface S sites are surrounded by significantly localized electron densities that are crucial for repulsive interactions in nc-AFM imaging. The density above the surface Sn atom is even more localized than that of the S atom, indicating the surface Sn atom behaving like an atom of a non-metallic element, which is in opposite to its usually metallic behavior as those bulk Sn atoms show. (**c**)-(**e**) Frequency-shift, damping, and current channels of an nc-AFM image. The green and orange triangles and the white hexagon represent the $α_{II}$, $β_{II}$, and $γ_{II}$ regions, respectively. The surface Sn atom represents the $α_{II}$ region (green triangle) in the SPM images. Its higher vertical position and the largest localized electron density leads to the strongest repulsion in the frequency shift image, exhibit no dissipation in the damping image and significant tunneling current in the STM image. As for the embedded Sn (the $γ_{II}$ region), it is highly metallic according to the cross-sectional ELF plot; this allows to dynamically form chemical binds between the embedded Sn and the O atom of the CO tip during scanning. As a result, significant damping signal (white hexagons in **d**) in a nearly hexagonal shape was recorded around the embedded Sn atoms while appreciable tunneling current was obtained in the associated STM image (white hexagons in **e**). The surface S atom (orange triangles, representing region $β_{II}$) sits at a position in between those two types of Sn atoms in terms of the vertical position, strength of Pauli repulsion and electronic conductance, so that it yields moderate Pauli repulsion, showing blurry triangles in the frequency shift image and nearly insulating feature in the STM image. Such assessment of Sn (surface), S and Sn (embedded) sites to regions $α_{II}$, $β_{II}$, and $γ_{II}$ is also consistent with the experimental measurement of force curves.

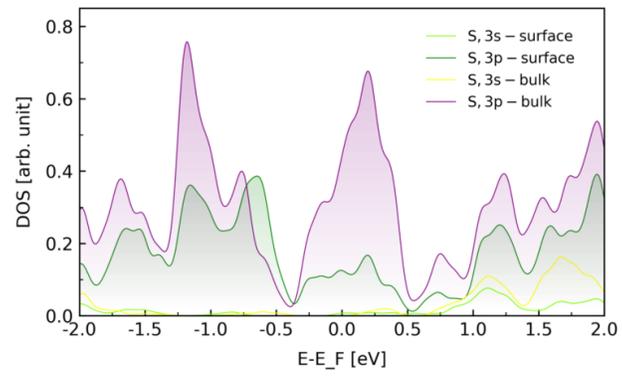

**Extended Data Fig. 5 | DFT calculated PLDOS of S orbitals in Sn terminated Co$_3$Sn$_2$S$_2$.** Between the energy range of 0.38-0.50 eV, only the *p* orbitals participate in the hybridization, both for the surface and bulk S atoms.

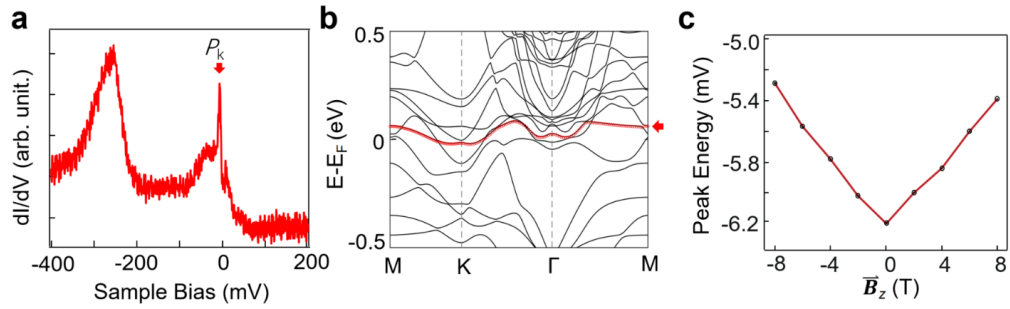

**Extended Data Fig. 6 | Physical properties of the SKES on Sn surface of $Co_3Sn_2S_2$.** (**a**) STS on Sn surface. The red arrow marks the flat-band related electronic state residing at -10 mV. ($V_s$= -400 mV, $I_t$ = 100 pA, $V_{mod}$ = 0.5 mV). (**b**) DFT calculated electronic band structure with the SOC included where the flat-band is highlighted in red. (**c**) Energy shift of the flat-band peak position as a function of vertical magnetic field, showing the feature of negative orbital magnetism.

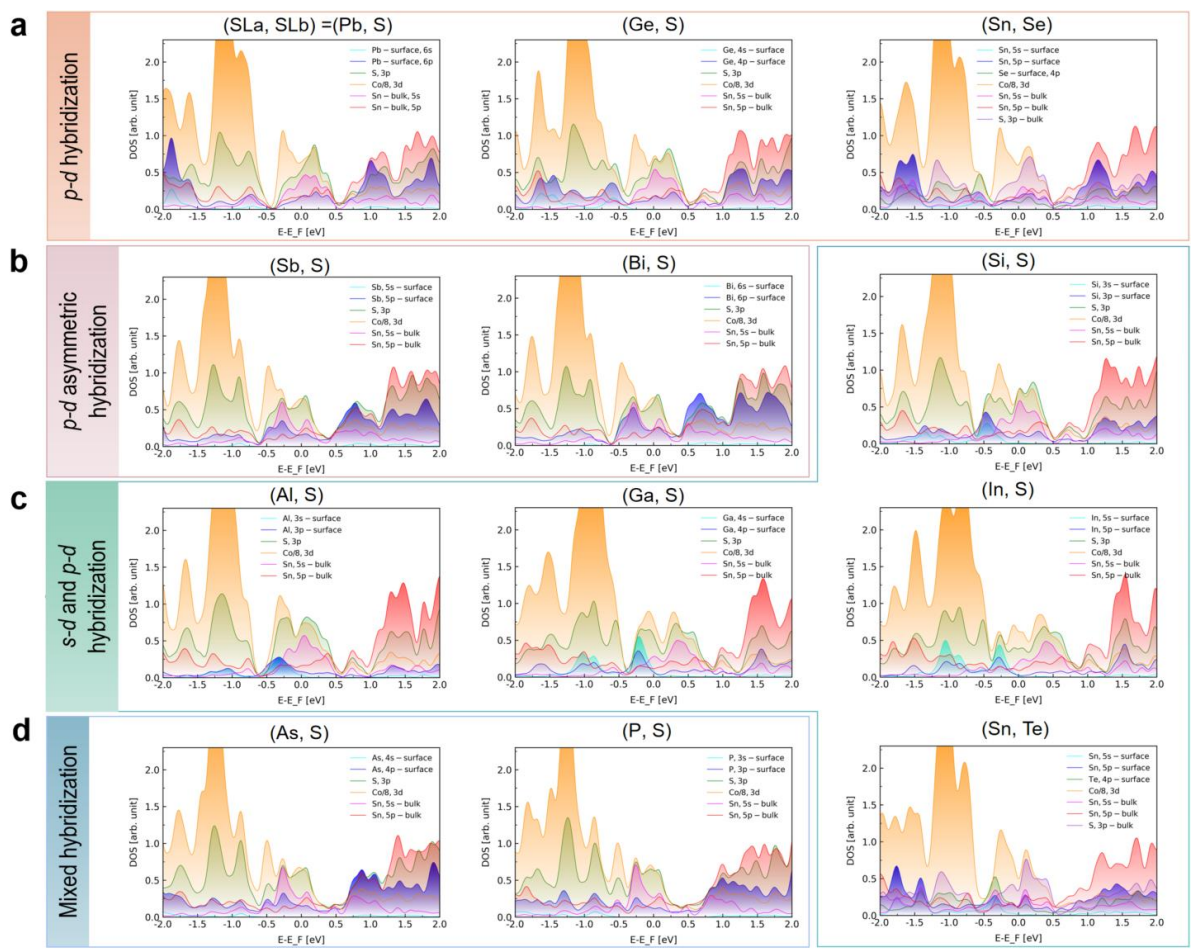

**Extended Data Fig. 7 | SKES constructed with different elements in SLa and SLb on Co$_3$Sn$_2$S$_2$.** (**a**) *p-d* hybridized surface structure by depositing Pb or Ge elements on S surface or substitute the subsurface S atoms with Se atoms on Sn surface. (**b**) Asymmetric *p-d* hybridized surface structure by depositing Sb or Bi elements on S surface. (**c**) *s-d* and *p-d* hybridized surface structure by depositing Al, Ga, In or Si elements on S surface, or substitute the subsurface S atoms with Te atoms on Sn surface. (**d**) Mixed hybridized surface structure by depositing As or P elements on S surface.